\documentclass[12pt,showpacs,preprintnumbers,amsmath,amssymb]{revtex4}
\usepackage{amsmath,bm,graphicx}
\usepackage{amssymb}
\usepackage{amsfonts}
\usepackage{color}
\begin{document}

\title{Simulations: the dark side}

\author{Daan Frenkel}
\email[]{df246@cam.ac.uk}
\affiliation{Department of Chemistry, University of Cambridge, Lensfield Road, Cambridge, CB2 1EW, United Kingdom}
 
 \date{\today}

\begin{abstract}
This paper discusses the Monte Carlo and Molecular Dynamics methods. 
Both methods are, in principle, simple. However,  simple does not mean risk-free. 
In the literature, many of the pitfalls in the field are mentioned,  but usually as a footnote -- and these footnotes are scattered over many papers.
The present paper focuses on the `dark side' of simulation: it is one big footnote. 
I should stress that `dark', in this context, has no negative moral implication. 
It just means: under-exposed. 
\end{abstract}
\maketitle
\section{Introduction and acknowledgment}
At the 2012 Varenna summer school on {\em Physics of Complex Colloids}, I gave a series of lectures on computer simulations in the context of complex liquids. 
The lectures were introductory, although occasionally, I would mix in a more general cautionary remark. 
It seemed to me that there was little point in writing a chapter in the Proceedings on `Introduction to Computer Simulations'. Books on the topic exist. 
However, I did not quite know what to write instead. Then, over lunch,  {\em Wilson Poon} suggested to me to write something on the limitations of existing simulations methods: where do they go wrong and why? I liked the idea very much.

The scope of the present manuscript is a bit broader: after a fairly general (but brief) introduction, I will discuss three types of issues:
\begin{enumerate}
\item Computer simulation methods that seem simple \ldots yet require great care
\item Computer simulation methods that seem reasonable \ldots but are not
\item Myths and misconceptions
\end{enumerate}
Not all issues that I list are of direct relevance for soft matter. However, I hope that the reader will forgive me. 
I should also point out that many of the issues that I discuss are very well known -- sometimes they are even trivial. 
However, I thought it better to list even the trivial examples, rather than assume that every single one of them is well known to all readers. 
Some of the issues that I highlight may not be well known, simply because I am mistaken or I have missed a key reference.
If so, I apologise. I also apologise for the rather black-or-white way in which I present problems. 
Seen in their original context, the issues are usually more subtle. 
My aim is to show what can go wrong if techniques are used  outside their original context. 

\section{Simulations: why and why not}
Over the past 60 years, the speed at which computers perform elementary calculations has increased by a factor 10$^{15}$, and the size of computer memories and the capacity of data storage devices have undergone similarly spectacular increases. The earliest computer simulations of systems consisting of a few hundred atoms could only be performed on the world's largest computers. Now, anybody who has access to a standard computer for personal use can carry out simulations that would have required a supercomputer only 15 years ago. Moreover, software to carry out computer simulations is readily available. The fact that the hardware and software thresholds for performing `normal' simulations have all but disappeared forces us to think about the role of computer simulations~\footnote{In the past, we had to think about the role of simulations because they were expensive, now we have to think because they are (mostly) cheap.}. The key question is: `Why should one perform a simulation in the first place.

\subsection{Predicting properties of novel compounds or materials}
When we look at computer simulations in an applied context, the answer to the question `why simulation?' is simple: they can save time (and money). 
Increasingly, simulations are used to complement experiment or, more precisely, to guide experiments in such a way that they can focus on the promising compounds or materials. This is the core of the rapidly growing field of computational materials science and computational `molecular' design. 
Computer simulations allow us to predict the properties of potentially useful substances, e.g. pharmaceutical compounds or materials with unique physical properties. Using computer simulations we can pre-screen candidate substances to minimise the amount of experimental work needed to find a substance that meets our requirements. In addition, simulations are very useful to predict the properties of materials under conditions that are difficult to achieve in controlled experiments (e.g. very high temperatures or pressures). 

Computational materials science of the type sketched above is the `front end' of a broader scientific endeavour that aims to advance the field of particle-based modelling, thus opening up new possibilities. Much of this development work is carried out in an academic environment where other criteria apply when we wish to answer the question whether a simulation serves a useful purpose. 
Below, I list several valid reasons to perform a simulation, but I also indicate what reasons I consider less convincing. Let me begin with the latter.
\subsection{Because it's there}
The total number of molecular systems that can, in principle, be simulated is very, very large. 
Hence, it is not difficult to find a system that nobody else has simulated before. This may seem very tempting. It is easy to perform a simulation, create a few nice colour snapshots and compute, say, a radial distribution function. Then, we write a manuscript for a high impact journal and, in the abstract, we write `Here, for the first time, we report Molecular Dynamics simulations of {\em 18-bromo-12-butyl-11-chloro-4,8-diethyl-5-hydroxy-15-methoxytricos-6,13-diene-19-yne-3,9-dione}' -- I took the name from Wikipedia, and my guess is that nobody has simulated this substance. Then, in the opening sentence of our manuscript we write: `Recently, there has been much interest in the Molecular Dynamics of {\em 18-bromo-12-butyl-11-chloro-4,8-diethyl-5-hydroxy-15-methoxytricos-6,13-diene-19-yne-3,9-dione ...}' and, with a few more sentences, a some snapshots and graphs, and a concluding section that mirrors the abstract, the work is done...

Of course, this example is a parody of reality -- but only just. Such simulations provide information that answers no existing question -- it is like the famous passage in the Hitchhikers Guide to the Galaxy, where the computer `Deep Thought' has completed a massive calculation to answer the question of {\em Life, the Universe and Everything}. The answer is 42 -- but the problem is that nobody really remembers what the question was. 

A simulation should answer a question. But there are different kinds of questions. I will discuss some of the categories below.
\subsection{Testing force fields}
Our knowledge of forces between all but the simplest molecules is limited. Moreover, the construction of reliable force-fields is a time-con\-su\-ming business that combines experimental and theoretical information. The most widely used force fields are hopefully `transferable'. This means that the interactions between molecules are decomposed into interactions between their constituent atoms or groups, and that the same interactions can be used when these atoms or groups are arranged to form other molecules. The model may take interactions between charges and polarisability into account but, in the end, the force-field is always approximate. This is even true when interactions are computed `on the fly' using density-functional theory or another quantum-based approach. 

Therefore, if we wish to apply a force field to a new class of molecules, or in a range of physical conditions  for which it has not been tested, we cannot assume {\em a priori} that our force-field will perform well: we must compare the predictions of our simulations with experiment. If simulation and experiment disagree, our force field will have to be improved. Optimising and validating force fields is an important ingredient of computer simulations.

\subsection{Simulation or theory}
It has been suggested that simulation constitutes a third branch of science, complementing experiment and theory. This is only partly true. There is a two-way relation between theory and experiment: theories make predictions for experimental observations and, conversely,  experiments can prove theories wrong -- if experiment agrees with theory then this does not demonstrate that the theory is correct but that it is compatible with the observations. Simulations have some characteristics of theory and some of experiment. 
\subsubsection{{\em Model testing}}
As is the case for theory, simulations start with the choice of a model, usually a choice for the Hamiltonian (`the force field') describing the system and, in the case of dynamical simulations, a choice for the dynamics (Newtonian, Langevin, Brownian etc). With this choice, simulations can (in principle) arrive at predictions of any desired accuracy -- the limiting factor is computing power. However, simulations can never arrive at exact results, let alone analytical relations. Unlike theories, simulations are therefore not a good tool to summarise our understanding of nature. However, they can provide important insights: for instance, a simulation can show that a particular model captures the essential physics that is needed to reproduce a given phenomenon. A case in point is hard-sphere freezing: before the simulations of Alder and Wainwright and Jacobson and Wood~\cite{AlderWainwright,WoodJacobson}, it was not obvious that systems with hard-sphere interactions alone could freeze. Once this was known, approximate theories of freezing could focus on the simple hard-sphere model -- and `computer experiments' could be used to test these theories. This shows two aspects of simulations: a) it can act as a `discovery tool' and b) it can be used to test predictions of approximate theories. 
\subsubsection{{\em The importance of exact results}}
Conversely, exact results from theory are an essential tool to test whether a particular algorithm works: if  simulation results (properly analysed and, if necessary, extrapolated to the thermodynamic limit) disagree with an exact theoretical result, then there is something wrong with the simulation. These tests need not be very sophisticated: they can be as simple as computing the average kinetic energy of a system and comparing with equipartition (for a system of the same size), or computing the limiting behaviour of the non-ideal part of the pressure of a system with short-ranged interactions at low  densities and comparing  with the value expected on the basis of our knowledge of the second virial coefficient.  
\subsubsection{{\em The digital microscope}}
When comparing simulation with experiment, we do something else: we are testing whether our Hamiltonian (`force-field') can reproduce the behaviour observed in experiment. As said before: agreement does not imply that the force field is correct, just that it is `good enough'. Once a simulation is able to reproduce the experimental behaviour of  a particular system, we can use the computer as a microscope: we can use it to find hints that help us identify the microscopic phenomena that are responsible for unexplained experimental behaviour.  
\subsubsection{{\em Simulation etiquette}}
\begin{figure}[ht]
\begin{center}
\includegraphics[width=0.8\textwidth]{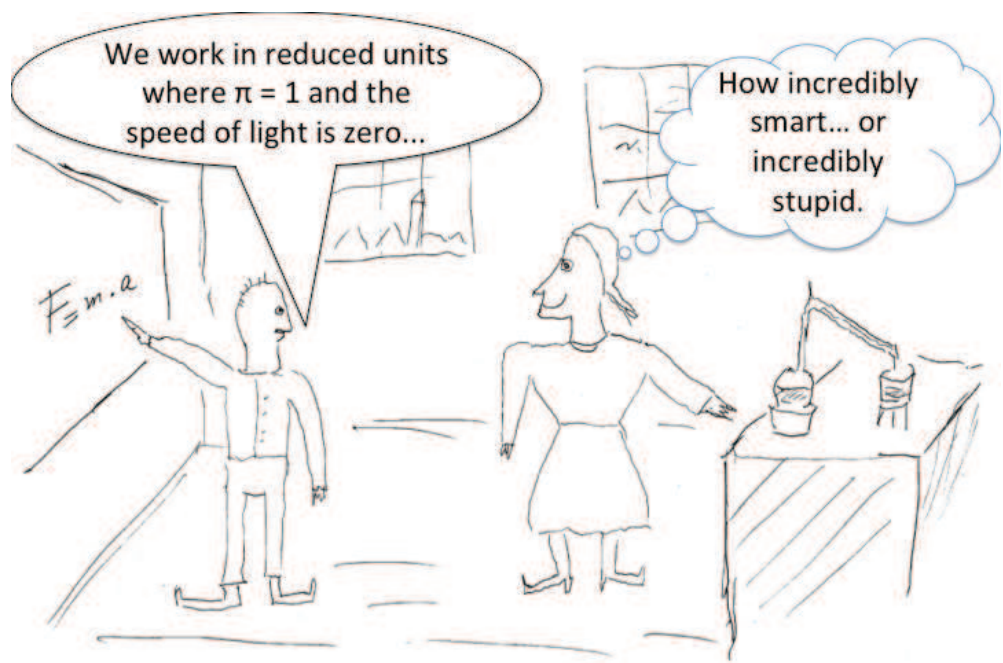} 
\caption{In simulations, it is common (and advisable) to express the primary simulation results in reduced units - be it not the ones in this figure. However, whenever comparing with experiment, the translation from reduced units to S.I. units should be made clear. Otherwise, confusion (or worse) will result.}
\label{fig:ReducedUnits}
\end{center}
\end{figure}

Experimentalists and simulators often speak a different `language'. Experimental data are usually (although not always) expressed in S.I. units. Simulators often use reduced units that can be an endless source of confusion for experimentalists (see Fig.~\ref{fig:ReducedUnits}). There are good reasons to use reduced units in simulations [e.g. to use a characteristic particle diameter as a unit of length and to use the thermal energy ($k_BT$) as a unit of energy]. However, the units should be clearly defined, such that a conversion to S.I. units can easily be made.
There is one general rule, or rather a form of numerical `etiquette', governing the comparison between experiment and simulation that is not often obeyed.
The rule is that, if at all possible, simulations should compute directly what is being measured in experiment. The reason is that it is often not straightforward to convert  experimental data into the quantities that are computed in a simulation: the procedure may involve additional approximations and, at the very least, is likely to lead to a loss of information. To give a specific example: scattering experiments do not measure the radial distribution function $g(r)$, rather, they measure $S(q)$ over a finite range of wave-vectors. As $g(r)$ is related to the Fourier transform of $S(q)$, we would need to know $S(q)$ for {\em all} wave vectors to compute $g(r)$. In practice, experiments probe $S(q)$ over a finite window of wave-vectors. If we would perform a Fourier transform within this window, we would obtain an estimate for $g(r)$ that contains unphysical oscillations (and possibly even negative regions), due to the truncation of the Fourier transform. Of course, procedures exist to mitigate the effects of such truncation errors, yet, if at all possible, simulations should report $S(q)$ to save the experimentalist the trouble of having to perform an approximate Fourier transform of their data.  There is, of course, a limit to this rule: some corrections to experimental data are specific for one particular piece of equipment. If simulations would account for these specific instrument effects, they could only be compared to one particular experiment. In that case, the meeting point should be half way.
\subsubsection{{\em Embedded simulation}}
Computers have become more powerful -- but so have experimental techniques (often for the same reason). Many experiments generate huge data streams that are impossible to analyse by hand. For this reason, computers play an important role in data analysis. Most of these techniques are not related to computer simulations -- but some are. Increasingly, molecular modelling is integrated in experiments. The experimental data provide constraints that allow us to refine the model. Increasingly, this type of simulation will be embedded in the experimental equipment -- a black box that most users will not even see. 

\subsection{Algorithms versus Moore's law}
The staggering increase in computing power during the past 60 years [roughly by a factor 10$^{15}$~\footnote{Of course, the increase is only partly due to increases in processor speed. Much of the recent gains come from increasing parallelism.}],  is such that it has not only quantitatively changed computing, but also qualitatively, because we can now contemplate simulations (e.g. a short simulation of a system the size of a bacterium) that were not even on the horizon in the early 50's. However, in some cases, the speed-up achieved through the use of better algorithms (e.g. techniques to study rare events) have been even more dramatic. Yet, the algorithms at the core of Monte Carlo or Molecular Dynamics simulations have barely changed since the moment that they were first implemented. Hence, a computer simulator in the 60's would have been able to predict fairly accurately on the basis of Moore's law, what system sizes and time-scales we can simulate today. However, he/she (more `he' than `she' in the 50's and  60's -- Arianna Rosenbluth and Mary-Ann Mansigh are notable exceptions) would not have been able to predict the tools that we now use to study rare events, quantum systems or free-energy landscapes. I should add that many algorithmic improvements are not primarily concerned with computing speed but with the fact that they enable new types of calculations (think of thermostats, barostats, fluctuating box shapes etc). It would be incorrect to measure the impact of these techniques in terms of a possible gain in computing speed. 

\subsection{When not to compute}
Although this point of view is not universally accepted, scientists are human.  Being human, they like to impress their peers. One way to impress your peers is to establish a record. It is for this reason that, year after year, there have been -- and will be -- claims of the demonstration of ever larger prime numbers: at present -- 2012 -- the record-holding  prime contains more than ten million digits but less than one hundred million digits. As the number of primes is infinite, that search will never end and any record is therefore likely to be overthrown in a relatively short time. No eternal fame there. 
In simulations, we see a similar effort: the simulation of the `largest' system yet, or the simulation for the longest time yet (it is necessarily `either-or'). 
Again, these records are short-lived. They may be useful to advertise the power of a new computer, but their scientific impact is usually limited. 
At the same time, there are many problems that cannot be solved with today's computers but that are tantalisingly close. It may then be tempting to start a massive simulation to crack such a problem before the competition do it. That is certainly a valid objective, and Alder and Wainwright's discovery of long-time tails in the velocity auto-correlation function of hard spheres is a case in point. The question is: how much computer time (`real' clock time) should we be prepared to spend. To make this estimate, let us assume that the project will have unlimited access to the most powerful computer available at the time that the project starts. Let us assume that this computer would complete the desired calculation in $\tau$ years. Somewhat unrealistically, I assume that the computer, once chosen, is not upgraded. The competition has decided to wait until a faster computer becomes available. We assume that the speed of new computers follows Moore's law. Then, after waiting for $t$ years, the time to completion on a new computer will be $\tau \exp(-t/2.17)$ (2.17 years is the time it takes the computer power to increase by a factor of $e$ if Moore's law holds). The question is: who will win, the group that starts first, or the group that waits for a faster computer to become available: {\em i.e.} is
\begin{equation}
\tau \le t+ \tau \exp(-t/2.17)\; .
\end{equation}
The time $t$ that minimises $ t+ \tau \exp(-t/2.17)$, satisfies:
\begin{equation}
\frac{2.17}{\tau}=\exp(-t/2.17)\; .
\end{equation}
This equation has a solution for positive $t$ when $\tau>2.17$ years. If the calculation takes less than 2.17 years, the group that starts first wins. Otherwise, the group that waits finishes the calculation after
\begin{equation}
\tau_c = 2.17 \left[\ln\left(\frac{\tau}{2.17}\right) +1\right]
\end{equation}
years.  Of course, this calculation is very oversimplified, but the general message is clear: it is not useful to buy a computer to do a calculation that will take several years. Once more, I should stress that I mean `wall clock' time -- not processor years.

Finally, it is important to realise that Moore's law is not a law but an extrapolation that, at some point, will fail. 
\section{Methods that seem simple\ldots but require great care}
\subsection{Finite  size effects}
Many of the subtleties of computer simulations are related to the fact that the systems that are being studied in a simulation are far from macroscopic. The behaviour of a macroscopic system can be imitated, but never really reproduced, by using periodic boundary conditions.  The main advantage of periodic boundary conditions is that they make it possible to simulate a small system that is not terminated by a surface, as it is periodically repeated in all directions.  But even though the use of periodic boundary conditions will eliminate surface effects (that are very dramatic for small systems), there are many other finite-size effects that cannot be eliminated quite so easily. Below, I will give a few examples. But first, a general comment about testing for finite size effects.
\subsubsection{{\em Half is easier than double}}
If a simulator suspects finite size effects in a simulation, the obvious test is to repeat the same simulation with a system of a larger size, e.g. a system with double the linear dimension of the original system. The problem with this test is that it is rather expensive. The computational cost of a simulation of a fixed duration scales (at least) linearly with the number of particles. For a 3D system, doubling the linear dimensions of the system makes the simulation 8 times more expensive. In contrast, halving the system size makes the simulation 8 times cheaper. Of course, if there is evidence for finite-size effects, then there it is necessary to perform a systematic study of the dependence of the simulation results -- and then there is no cheap solution. A systematic analysis of finite size affects if of particular importance in the study of critical phenomena~\cite{Binder}. In general, finite size effects will show whenever we study phenomena that are correlated over large distances: these can be static correlations (as in the case of critical fluctuations) or dynamic correlations, as in the case of hydrodynamic interactions.
\subsection{Hydrodynamic interactions}
When a particle moves in a fluid, it gradually imparts its original momentum to the surrounding medium. As momentum is conserved, this momentum spreads in space, partly as a sound wave, partly as an (overdamped) shear waves. The sound wave moves with the speed of sound $c_s$ and will therefore cross a simulation box with diameter $L$ in a time $L/c_s$. After that time, a particle will `feel' the effect of the sound waves emitted by its nearest periodic images. Clearly, this is finite size effect that would not be observed in a bulk fluid. The time $\tau_s$ for such spurious correlations to arise is, in practice, relatively short. If $L$ is of the order of 10~nm and $c_s$ is of the order $10^3$ m/s, then $\tau_s$ is of order 10 ps. Needless to say that most modern simulation runs are much longer than that. In addition to a sound wave, a moving particle also induces shear flow that carries away the remainder of its momentum. This `transverse' momentum is transported diffusively. The time to cross the periodic box is of order $L^2/\nu$, where $\nu$ is the kinematic viscosity ($\nu\equiv \eta/\rho_m$, where $\eta$ is the shear viscosity and $\rho_m$ the mass density). To take water as an example: $\nu=10^{-6}{\rm m^2 s^{-1}}$ and hence the time $\tau_\perp$ that it takes transverse momentum to diffuse one box diameter (again, I will assume $L$ = 10~nm) is around 10 ps. What these examples show is that any numerical study of  long-time dynamical correlations  may require a careful study of hydrodynamic finite-size effects. Below, I will discuss the specific example of hydrodynamic interactions. However, many of my comments apply (suitably modified) to the study of dislocations or the calculation of the properties of systems consisting of particles that interact through long-ranged forces (e.g. charged or dipolar particles). 
\subsubsection{{\em Self diffusion coefficients}}
At large distances, the hydrodynamic flow field that is induced by a force acting on a particle in a fluid, decays as $r^{-1}$. If we now consider a system with periodic boundary conditions, the hydrodynamic flow fields due to all periodic images of the particle that is being dragged by the external force will have to be added. It is clear that, just as in the case of the evaluation of Coulomb interactions,  a naive addition will not work. It would even seem that the total flow field due to an infinite periodic array of dragged particles would diverge. This is not the case, and the reason is the same a in the Coulomb case: when we compute Coulomb interactions in a periodic system, we must impose charge neutrality. Similarly, in the case of hydrodynamic interaction, we must impose that the total momentum of the fluid is fixed (in practice: zero). This means that if we impart a momentum ${\bf p}$ to a given article, then we must distribute a momentum $-{\bf p}$ among all other particles. The result is that, if we start with a fluid at rest, dragging one particle in the $+x$ direction, creates a back flow of all other particles in the $-x$ direction. If there are $N$ particles in the system, this is an ${\mathcal O}(N^{-1})$ effect. But that is not the only finite size effect. The other effect is due to the fact that the drag force per particle needed to move  a periodic array of particles through a fluid is not simply the sum of the forces needed to move a single particle at infinite dilution. In fact, the force is different (larger) by an amount that is proportional to $L^{-1}$ (roughly speaking, because the hydrodynamic interactions decay as $r^{-1}$). Hence, the mobility (velocity divided by force) of a particle that belongs to a periodic array is less than the of an isolated particle in a bulk fluid. As a consequence (Stokes-Einstein), the diffusion coefficient of a particle in a periodic fluid has a substantial finite-size correction that scales as $\sigma/L$, where $\sigma$ is the diameter of the particle. For a three-dimensional system, the diffusion coefficient therefore approaches its infinite system-size limit as $N^{-1/3}$, {\em i.e.} very slowly. By comparison, the finite-size correction due to back flow is relatively minor.
\subsection{Compatibility of boundary conditions}
Periodic boundary conditions impose a specific symmetry on the system under study. For instance, if we study a fluid using simple-cubic periodic boundary conditions, then a single `cell' of this periodic lattice may look totally disordered, yet as it is repeated periodically in all directions, the system should be viewed as a simple cubic crystal with a particularly complex unit cell. As was already well known in the early days of computer simulations, the imposed symmetry of the periodic boundary conditions induces some degree of bond-orientational order in the fluid~\cite{Mandell}. Often, this (slight) order has no noticeable effect, but in the case of crystal nucleation it may strongly enhance the nucleation rate. This is why finite size effects on crystal nucleation are so pronounced.

The shape of the simulation box has an effect on the magnitude of finite size effects. Intuitively, it is obvious that this should be the case in the limit that one dimension of the box is much smaller than the others. For instance, if the simulation box has the shape of a rectangular prism with sides $L_x\ll L_y\le L_z$, then the shortest distance between a particle and its periodic image is $L_x$. Clearly, we want $L_x$ to be sufficiently large that a particle does not `feel' the direct effect of its periodic images. This sets a lower limit for $L_x$. For a cubic box, $L_y$ and $L_z$ would have the same value, but for the rectangular prism mentioned above, they will be larger. Hence, a large system volume would be needed to suppress finite size effect for a rectangular prismatic box than for a cubic box. This is one of the reasons why cubic boxes are popular; the other reason is that it is very easy to code the periodic boundary conditions for cubic boxes. This second reason explains why cubic boxes are being used at all, because there are other box shapes that result in a larger distance between nearest periodic images for the same box volume. Depending on the criteria that one use, there are two shapes that can claim to be optimal. The first is the {\em rhombic dodecahedron} (i.e. the Wigner-Seitz cell of a face-centred cubic lattice). For a given volume, the rhombic dodecahedron has the largest distance between nearest periodic images -- or, what amounts to the same thing, it has the largest inscribed sphere. The other (more popular) shape is the {\em truncated octahedron} (the  Wigner-Seitz cell of a body-centred cubic lattice). The truncated octahedron is the most `spherical' unit cell that can be used, in the sense that it has the smallest enclosing sphere. In fact, the ratio of the radii of the enclosing and inscribed spheres of the truncated octahedron is $\sqrt{5}/3\approx 1.29$. For a rhombic dodecahedron, this ratio is $\sqrt{2}\approx 1.41$ and for a cube it is $\sqrt{3}\approx 1.73$. In two dimensions the optimal box shape is a hexagon (and in 4D it is the Wigner-Seitz cell of the so-called  $D_4$ lattice). 

In simulations of isotropic fluids, any shape of the simulation box can be chosen, provided that it is sufficiently large to make finite-size effects unimportant.
However, for ordered phases such as crystals, the choice of the simulation box is constrained by the fact that is has to accommodate an integral number of crystal unit cells. If the crystal under study is cubic, then a cubic simulation box will do, otherwise the simulation box should be chosen such that it is as `spherical' as possible, yet commensurate with the unit cell of the crystal under study. Note that this constraint implies that, if a given substance has several crystal polymorphs of lower than cubic symmetry, then different box shapes will be needed to simulate these phases. 
Moreover, for low symmetry crystals, the shape of the crystal unit cell will,  in general, depend on temperature and pressure. Hence, simulations of different thermodynamic state points will require simulation boxes with different shapes. If the shape of the simulation box is incompatible with the crystal symmetry, the crystal will be strained. This is easy to verify by computing the stress tensor of the system: if the average stress is not isotropic (hydrostatic), the shape of the simulation box causes a deformation of the crystal. As a consequence, elastic free energy will be stored in the crystal and it will be less stable than  in its undeformed state.  Not surprisingly, such a deformation will affect the location of phase coexistence curves. The effect of the shape of the simulation box on the crystal under study is not simply a finite size effect: even a macroscopic crystals  can be deformed. However, for a large enough system, it is always possible to choose the number of crystal unit cells in the $x,y$ and $z$ directions such that the resulting crystal is almost commensurate with the simulation box. The remaining difference is a finite size effect. However, if we have been less careful and have prepared a deformed crystal, then it will stay deformed, if not forever, then at least much longer than the time that we can study in a simulation.

\subsubsection{{\em Deformable boxes}}
In practice, the shape of the simulation box is not chosen by trial and error until the crystal is stress free (or, more precisely, has an isotropic stress).  In the early 80's, Parrinello and Rahman~\cite{ParrinelloRahman} developed a constant-stress Molecular Dynamics scheme that treated the parameters characterising the box shape as dynamical variables. In a simulation, the box shape fluctuates and the average box shape is the one compatible with the imposed stress. If the imposed stress is isotropic, then the Parrinello-Rahman technique yields the box shape compatible with an undeformed crystal. 
Shortly after its introduction,  the Parrinello-Rahman technique was extended to Monte Carlo simulations by Najafabadi and Yip~\cite{NajafabadiYip}. 
More recently, the method of Najafabadi and Yip was used by Filion and coworkers~\cite{FilionDijkstra} to predict the crystal structure of novel materials. In this method, small systems are used such that appreciable fluctuations in the shape of the simulation box are possible: this allows one to explore a wide range of possible structures. 

Potentially, the fluctuations in the shape of the simulation box can become a problem: if the box becomes very anisometric, the nearest-neighbour distance in some directions becomes very small and serious finite size effects are observed. This problem is always present if fluctuating box shapes are used to simulated liquids (but, usually, there is no need to use fluctuating box shapes in that case). In the case of crystals, the problems are most pronounced for small systems. In that case, the fluctuations of the box shape may have to be constrained. To allow for larger changes in the shape of the crystal unit cell, the system should be mapped onto a new simulation box, once the original box becomes too deformed. This remapping does not affect the atomic configuration of the system.

\subsection{Liquid crystals}
Liquid crystals are phases with symmetry properties intermediate between those of an isotropic liquid and of a fully ordered, three-dimensional crystal.
The simplest liquid crystal, the so-called {\em nematic} phase, has no long-range translational order, but the molecules are on average aligned along a fixed direction, called the nematic {\em director}. Like isotropic liquids, nematic phases are compatible with any shape of the simulation box.
However, for more highly ordered liquid-crystalline phases, the box shape should be chosen carefully. To give a simple example: the smectic-A phases is similar to the nematic phase in that all molecules are, on average aligned along a director but, unlike the nematic phase, the density of the nematic phase along the director is periodic.  In the simplest case one can view the smectic-A phase as a stack of liquid-like molecular layers. Clearly, in a simulation, the boundary conditions should be such that an integral number of layers fits in the simulation box. This constrains the system size in one direction (viz. along the director), but not along the other two directions. Again, simulations with a flexible box shape can be used to let the system relax to its equilibrium layer spacing. However, as the smectic phase is liquid like in the transverse direction, one should make sure that the box shape in those directions is fixed. 
\subsection{Helical boundary conditions} 
\begin{figure}[h!]
\begin{center}
\includegraphics[width=0.8\textwidth]{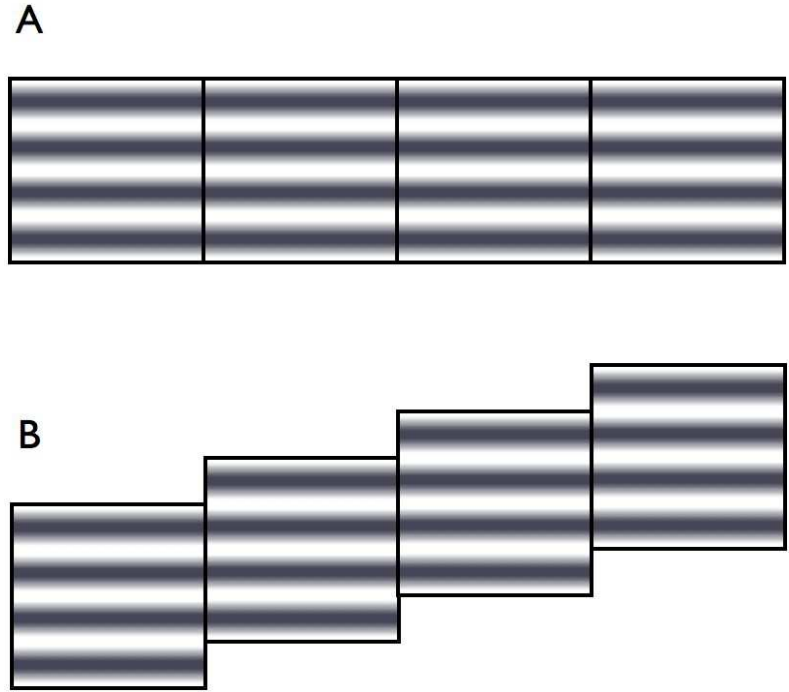} 
\caption{(A): With normal boundary conditions the system contains four distinct smectic layers. (B): With helical boundary conditions, these four layers are all connected and can therefore easily exchange particles.}
\label{fig:helical}
\end{center}
\end{figure}

There is a practical reason for using slightly more complex boundary conditions for smectic-A phases: the average areal density in all smectic-A layers is the same, but the number of molecules per layer will fluctuate. This is a real effect. However, the time it takes a molecule to diffuse (`permeate') from one smectic layer to the next tends to be very long (it involves crossing a free-energy barrier). Hence, unless special precautions are taken, the distribution of molecules over the various smectic layers will hardly vary over the course of a simulation and hence, the system does not adequately sample the configuration space. This problem can be solved by using so-called `helical' boundary conditions. Suppose that we have a smectic phase in a rectangular prismatic simulation box. We assume that the smectic layers are perpendicular to the $z$ axis and have an equilibrium spacing $d$. Now consider the next periodic box in the $x$-direction. Normally, this box is placed such that the point $x,y,z$ maps onto $x+L_x, y,z$. If we use helical boundary conditions, the box at $L_x$ is shifted by $d$ along the $z$ direction: hence, $x,y,z$ maps onto $x+L_x, y,z-d$ (see Fig.~\ref{fig:helical}). 

The result is that all the smectic layers in the system are now connected laterally into a single layer. Now, fluctuations in the number of particles per layer is easy because it does not require permeation. Similar techniques can be used to allow faster equilibration of columns in the so-called {\em columnar} phase (a phase that has crystalline order in two directions and liquid-like order in the third). 

I should point out that helical boundary conditions are not necessarily more `realistic' than normal periodic boundary conditions. In particular, the slow permeation of particles between layers is a real effect. 

\subsubsection{{\em Twist}}
The simulation of cholesteric liquid crystals poses a special problem. A cholesteric liquid crystal is similar to the nematic phase in that it has no long-ranged translational order. However, whereas the director if a nematic phase has a fixed orientation throughout the system, the director of a cholesteric phase varies periodically (`twists') in one direction and is constant in any plane perpendicular to the twist direction. For instance, if we choose the twist direction to be parallel to the $z$-axis, then the $z$-dependence of the director  would be of the form
\begin{equation}
\begin{array}{lcl}
n_z (z) & = & \mbox{constant}\; ,\\
n_x(z) & = & n_x(0)\cos(2\pi z/{\mathcal P})\; , \\
n_y(z) & = & n_y(0)\sin(2\pi z/{\mathcal P})\; , \\
\end{array}
\end{equation}
where ${\mathcal P}$ denotes the cholesteric pitch. Ideally, the periodic boundary conditions should be chosen such that the cholesteric pitch is not distorted. In practice, this is often not feasible because typical cholesteric pitches are very long compared to the molecular dimensions. For cholesteric phases of small molecules (as opposed to colloids), pitches of hundreds of nanometers are common. Simulations of systems with such dimensions would be very  expensive. The problem can be alleviated a bit by using twisted periodic boundary conditions. In that case the system is not simply repeated periodically in the $z$-direction (the pitch axis), but it is rotated by an amount that is compatible with the boundary conditions~\cite{AllenMasters}. If the simulation box is a square prism, any rotation by a multiple of $\pi/2$ is allowed. The minimal twist corresponds to the case where the rotation is $\pm \pi/2$. If the simulation box is a hexagonal prism, rotations that are a multiple of $\pi/3$ (corresponding to 1/6-th of a pitch) are allowed. Smaller angles are not possible.  

Although ${\mathcal P}/6<{\mathcal P}$, the box size needed to accommodate an undeformed cholesteric is usually still too large to be computationally feasible. Hence, cholesterics are usually simulated in an over- or under-twisted geometry and the properties of the undistorted are deduced using thermodynamics. Allen~\cite{Allen} has used such an approach  to compute the twisting power of chiral additives in a nematic phase.

Here, I outline how a simulation of  over- and under-twisted cholesterics can be used to estimate the equilibrium pitch. 
To do so, we have to make use of the fact that, for small deformations,  the free energy of a cholesteric phase depends quadratically on the degree of over(under)-twist~\footnote{It is not a priori obvious that, for the large twists that are applied in simulations, the quadratic approximation holds.}:
\begin{equation}
\Delta F(\xi)\approx \frac{V}{2}K_2 (\xi-\xi_0)^2\; ,
\end{equation}
where $\xi=2\pi/{\mathcal P}$ and $\xi_0$ is the value of $\xi$ for which the system is torque-free and  $V$ is the volume of the system. The constant $K_2$ is the so-called `twist elastic constant'. The subscript $2$ is there for historical reasons (in a nematic, two other types of director deformation are possible: `splay' and `bend': the associated elastic constants are denoted by $K_1$ and $K_3$ respectively). In what follows, I drop the subscript of $K_2$. In a periodic system with a box length $L_z$ in the $z$-direction,  $\xi \equiv \Delta\phi/L_z$, where  $\Delta\phi$ is the twist of the system over a distance $L_z$. In practice, we will only consider $\Delta\phi=0$ or $\Delta\phi=
\pm \pi/2$. Then an infinitesimal change in the Helmholtz free energy $F$ is given by
\begin{equation}
dF = -SdT -PdV+VK(\xi-\xi_0)d\xi+\mu dN\;.
\end{equation}
where $S$ is the entropy of the system, $T$ the temperature, $P$ the pressure, $\mu$ the chemical potential (per particle) and $N$ the number of particles.   In what follows, I fix $N$ and $T$. It is useful to write $dV$ as the sum of a volume change due to $dV_\perp$, a lateral expansion/compression of the system
 and $dV_\parallel= L_xL_y dL_z$, the volume change due to deformations in the $z$-direction. Note that the total free energy change due to a change in $L_z$ is given by
 \begin{equation}
 dF= \left[-PL_xL_y - VK(\xi-\xi_0)\frac{\Delta\phi}{L_z^2}\right]dL_z\;.
 \end{equation}
 We can write this as
  \begin{equation}
 dF_\parallel= -\left[P + K(\xi-\xi_0)\frac{\Delta\phi}{L_z}\right]dV_\parallel= -\left[P + K(\xi-\xi_0)\xi\right]dV_\parallel\;.
 \end{equation}
A change in volume at constant pitch (i.e. constant $L_z$) would yield
  \begin{equation}
 dF_\perp= -PdV_\perp\;.
 \end{equation}
 The excess pressure for deformations along the $z$-axis is therefore
 \begin{equation}
 \Delta P\equiv P_\parallel-P_\perp = K(\xi-\xi_0)\xi \;.
 \end{equation}

By measuring 
\begin{equation}
\frac{ \Delta P(\xi)-\Delta P(-\xi)}{\Delta P(\xi)+\Delta P(-\xi)} \approx \frac{ \xi_0}{\xi}\;,
\end{equation}
we can determine $\xi_0$. 

\subsection{Conserved quantities}
Momentum conservation in classical mechanics follows if the Lagrangian of the system is invariant under infinitesimal translations. This result carries over from a free system in the absence of external forces to a system with periodic boundary conditions. For a non-periodic system, momentum conservation implies that the centre of mass of the system moves at a constant velocity. In the presence of periodic boundary conditions,  this statement becomes ambiguous because one should specify how the position of the centre of mass is computed. Clearly if we interpret the centre of mass of the system as the mass-weighted average position of the particles in the simulation box at the origin, then the center of mass motion is not at all conserved. Every time the particle crosses the boundary of the simulation box -- and is put back at the other side -- the centre of mass undergoes a jump. However, if we consider the center of mass of the particles that were originally in the `central' simulation box, and do not put these particles back in that box when they move to another box, then the velocity of the center of mass of those particles is conserved.

It should be stressed that the very concept of the `boundary' of the simulation box is confusing: for normal (equilibrium) simulations, we can draw the lattice of simulation boxes wherever we like, hence the boundaries between such boxes can be anywhere in the system. The `boundary' therefore has no physical meaning (the situation is different in non-equilibrium simulations with so-called Lees-Edwards `sliding' boundary conditions).

Unlike linear momentum, angular momentum is not a conserved quantity in systems with periodic boundary conditions. The basic reason is that the Lagrangian is not invariant under infinitesimal rotations. This invariance (for spatially isotropic systems) is at the origin of angular momentum conservation. In periodic systems, angular momentum is simply not uniquely defined. To see this, consider a system consisting of two particles zero net linear momentum. In a free system, the angular momentum of this system is uniquely defined as it does not depend on the origin from which we measure the angular momentum. However, it is easy to see that in a periodic system the apparent angular momentum will be different if we choose the origin of our coordinate system between the two particles in the same simulation box, or between a particle in one simulation box and the periodic image of the other particle. Considering only particles in the same simulation box will not help because then the angular momentum will change discontinuously whenever we move a particle that has crossed a boundary back into the original simulation box.

\subsection{Computing S(q)}
There is another, more subtle, effect of periodic boundary conditions that is not always properly realised. It has to do with the calculation of Fourier transform of spatial correlation function. As an example (and an important one at that), I consider the computation of the structure factor $S(q)$. For simplicity, I will consider an atomic fluid. 
$S(q)$ is related to the mean square value of $\rho(q)$, the Fourier component of the number density at wave vector $q$:
\begin{equation}\label{eq:rhoq2}
S(q)=\frac{1}{N}\langle|\rho(q)|^2\rangle \;,
\end{equation}
where
\begin{equation}
\rho(q)\equiv \sum_{i=1}^N \exp\left(-\imath {\bf q\cdot r}_i\right)\; .
\end{equation}
Note that the structure factor is, by construction, non-negative. 
In liquid state theory, we often use the fact that the structure factor of a fluid is related to the radial distribution function $g(r)$ though
\begin{equation}
S(q)=1+\rho\int_V d{\bf r}\; \exp\left(-\imath{\bf q\cdot r}\right)\left[g(r)-1\right]\; .
\end{equation}
In the thermodynamic limit, the volume of the system tends to infinity and the integral does not depend on the size of the integration volume.
However, in a small, periodic system it does. More importantly, in an isotropic fluid, the radial distribution function is isotropic and we can write
\begin{equation}\label{eq:intgr}
S(q)=1+\rho\int_0^\infty 4\pi\; r^2 dr\; \frac{\sin qr}{qr}\left[g(r)-1\right]\; .
\end{equation}
For a finite, periodic system, one might be tempted to replace the upper limit of this integration by half the box diameter.
However, this is very dangerous because the function thus computed is no longer equivalent to the expression in Eq.~(\ref{eq:rhoq2}). In particular, the apparent $S(q)$ is not guaranteed to be non-negative, even if one only considered those values of $q$ that are compatible with the periodic boundary conditions. One solution to this problem is to extrapolate $g(r)$ to larger distances, using a plausible approximation. However, in general, it is safest to stick with Eq.~(\ref{eq:rhoq2}), even though it is computationally more demanding. 

\subsection{Using direct coexistence simulations to locate the freezing transition}
Computer simulations provide a powerful tool to locate first-order phase transitions. 
However, it is important to account properly for finite size effects. The safest (although not necessarily simplest) way to locate first order phase transitions
is to compute the free energy of the relevant phases in the vicinity of the coexistence curve. For a given temperature, this involves computing the free energy of both phases at one (or a few) state points and then computing the pressure  of the system in the vicinity of these points. In this way, one can evaluate the free-energy as a function of volume. Coexistence is then determined using a double tangent construction. The advantage of this procedure is that finite-size effects are relatively small as the two phases are simulated under conditions where no interfaces are present. However, the approach is rather different from the one followed in experiment where one typically looks for the point where the two phases are in direct contact. If the interface between the two phases does not move, we have reached coexistence. 

There are several reasons why, in simulations, this approach requires great care. First of all, the creation of a two-phase system involves creating a slab of phase I in contact with a slab of phase II (the slabs are obviously periodically repeated).  The free energy of such a system is equal to the free energy of the coexisting bulk phases plus the total surface free energy. A (potential) problem is that, if the slabs are not thick enough (how thick depends on the range of the intermolecular potential and on the range of density correlations), the total surface free energy is not equal to that of two surfaces at infinite separation. A second problem, that is more relevant for Monte Carlo than for Molecular Dynamics simulations is that two-phase systems equilibrate slowly. In particular, coexistence between two phases requires that the pressures are equal. In Molecular Dynamics simulations, the rate of pressure equilibration is determined by the speed of sound -- typically such equilibration is rapid. In contrast, in Monte Carlo simulations, pressure equilibrates through diffusion, and this is slower. 

Most likely, our first guess for the coexistence pressure at a given temperature will not be correct and we will see the phase boundary moving. Then we have to adjust the pressure. However, it is very important that we do not impose an isotropic pressure on the system. Rather, we should vary the volume of the system by changing the box-length in the direction perpendicular to the interface. The reason is that, in the transverse direction, the surface tension of the two interfaces contributes to the apparent pressure. Hence if the `longitudinal' pressure (i.e. the one acting on a plane parallel to the interfaces) is denoted by $P_\parallel$, then the apparent pressure in the perpendicular direction is 
\begin{equation}\label{eq:DeltaPgamma}
P_\perp= P_\parallel - \frac{2\gamma}{L_\parallel} \;.
\end{equation}
In other words: the transverse pressure contains a contribution due to the Laplace pressure of the interfaces. The coexistence pressure is $P_\parallel$, not $P_\perp$.

In the above example, we have considered the case of a liquid-liquid interface where the Laplace correction is determined by the surface tension $\gamma$. However,  when  one of the coexisting phases is a crystalline solid the problems are more severe because the pressure inside a solid need not be isotropic. In that case, the procedure becomes more complicated. First, one must determine the compressibility of the crystal or, more generally,  the relation between the lattice constant of the bulk crystal and the applied isotropic pressure. Then the lateral dimensions of the crystalline slab are chosen such that it is commensurate with a crystal with the correct lattice constant for the applied longitudinal pressure. If the initial guess for the coexistence pressure was incorrect one should change the applied longitudinal pressure -- but also the lateral dimensions of the simulation box, such that the crystal again has the correct lattice constant corresponding to the applied pressure. The rescaling of the lateral dimension of the simulation box should also be performed if we do not fix the longitudinal pressure but $L_\parallel$. In that case we do not know a priori what the coexistence pressure will be, but we still must rescale the transverse dimensions of the simulation box to make sure that the stress in the bulk of the crystal is isotropic. 

Finally, there is a general point that, for a small system, the free energy cost to create two interfaces is not negligible compared to the bulk free energy. For sufficiently narrow slabs the result may be that the system will form a strained homogeneous phase instead of two coexisting phases separated by interfaces. 

In general, simulations that study phase equilibria by preparing two phases separated by interfaces will use biasing techniques to facilitate the formation of a two-phase system. For examples, see Refs.~\cite{BinderGamma, Angioletti}.
\subsection{Computing the surface free energy of crystals}
Eq.~(\ref{eq:DeltaPgamma}) illustrates the surface tension of a liquid-liquid interface can create a difference between the longitudinal and transverse components of the pressure that can be measured in a simulation. In principle, Eq.~(\ref{eq:DeltaPgamma}) can be used to determine the surface tension $\gamma$. However, Eq.~(\ref{eq:DeltaPgamma}) is only valid for the interface between two disordered phases (e.g. liquid-liquid or liquid-vapour). If one of the phases involved is a crystal, the situation becomes more complicated. In particular, the Laplace pressure is not determined by the surface tension (which, for a solid-liquid interface is called the `surface free-energy density'), but by the surface stress. To see why this is so, consider the expression of the surface free energy of an interface with area $A$:
\begin{equation}
F_s(A)=\gamma A\; .
\end{equation}
The surface stress $\sigma_S$ is the derivative of the surface free energy with respect to surface area:
\begin{equation}
\sigma_S=\frac{\partial \gamma A}{\partial A}= \gamma + A\frac{\partial \gamma}{\partial A}.
\end{equation}
For a liquid-liquid interface, $\gamma$ does not depend on the surface area $A$, as the structure of the surface remains the same when the surface is stretched. Hence, in that case $\sigma_S=\gamma$. However, for a solid, the structure of the surface is changed when the surface is stretched and hence $\gamma$ changes. In that case, $\sigma_S\ne\gamma$. As a consequence it is much more difficult to determine $\gamma$ for a solid-liquid interface than for a liquid-liquid interface. In fact, whereas for a liquid-liquid interface $\gamma$ can be obtained from Eq.~(\ref{eq:DeltaPgamma}), the determination of $\gamma$ for a solid-liquid interface requires a free-energy calculation that yields the reversible work needed to create an interface between two coexisting phases. An example of such a calculation can be found in Ref.~\cite{Angioletti}. 

\subsection{Planck's constant in classical simulations}
Classical mechanics and classical statistical mechanics are at the basis of a large fraction of all Molecular Dynamics and Monte Carlo simulations.
This simple observation leads to a trivial conclusion: the results of purely classical simulations can {\em never} depend on the value of Planck constant
And, indeed, the expressions that are used to compute the internal energy, pressure, heat capacity or, for that matter, the viscosity of a system are functions of the classical coordinates and momenta only. 

However, there seems to be an exception: the chemical potential, $\mu$. When we perform a grand-canonical simulation, we fix $\mu$, $V$ and $T$ and compute (for instance) the average density $\rho$ of the system. $\rho$, $V$ and $T$ do not depend on $h$, but $\mu$ does. To be more precise, we can always write $\mu$ as:
\begin{equation}
\mu=\mu^{ideal}+\mu^{excess}\; ,
\end{equation}
where $\mu^{excess}$ is a purely classical quantity that depends on the interactions between molecules. However, $\mu^{ideal}$ depends on Planck constant. For instance, for an atomic system, 
\begin{equation}
\mu^{ideal}=k_BT\ln \left(\rho\Lambda^3\right)\; ,
\end{equation}
where $\Lambda$ is the thermal de Broglie wavelength:
\begin{equation}
\Lambda=\left(\frac{h^2}{2\pi mk_BT}\right)^{1/2}\; .
\end{equation}
For a molecular system, the expression for $\mu^{ideal}$ is of a similar form:
\begin{equation}
\mu^{ideal}=k_BT\ln \left(\rho\Lambda^3/q^{int}\right)\; ,
\end{equation}
where $q^{int}$ is the part of the molecular partition function associated with the vibrations and rotations (and, possibly, electronic excitations) of an isolated molecule. $q^{int}$ is a function of $T$  -- but its value  depends on $h$. 

It would then seem that the results of classical simulations can depend on $h$. In fact, they do not. 
First of all, if we consider a system at constant temperature, the factors involving $h$ just add a constant to $\mu$. Hence, the value of $h$ will not affect the location of any phase transitions: at the transition, the chemical potentials in both phases must be equal, and this equality is maintained if all chemical potentials are shifted by a constant amount.

One might object that the density of a system itself depends on $\mu$ and therefore on $h$. This is true, but in reality it is not a physical dependence: a shift the absolute value of the chemical potential is not an observable property. One should recall that the chemical potential describes the state of a system in contact with an infinite particle reservoir at a given temperature and density. In experiments, such a reservoir can be approximated by a large system, but in simulations it is most convenient to view this reservoir as consisting of the same atoms or molecules, but with all inter-molecular interaction switched off.

To see this, consider two systems at temperature $T$: a system with $N$ particles in volume $V$ and a reservoir with $M-N$ particles in volume $V_0$. We assume that $M\gg N$. The density in the reservoir $\rho_0 = (M-N)/V_0\approx M/V_0$. The reservoir and the system can exchange particles. 
We now assume that the particle in the reservoir do not interact. To be more precise: we assume that all {\em inter}-molecular interaction have been switched off in the reservoir, but all {\em intra}-molecular interactions are still present. This ideal gas is our reference system.

It is straightforward to write down the expression for the partition function of the reservoir for a given number of particles (say $M-N$):
\begin{equation}
Q_{res}(M-N, V_0,T)=\frac{V_0^{M-N}(q^{int})^{M-N} }{(M-N)!\Lambda^{3(M-N)}} \; .
\end{equation}
The total partition function (system plus reservoir) is:
 \begin{equation}
 Q_{tot}=\sum_{N=0}^M \frac{V_0^{M-N} (q^{int})^{M-N}}{(M-N)! \Lambda^{3(M-N)} } \frac{ (q^{int})^{N}\int d{\bf r}^N\; \exp\left(-\beta U(r^N)\right)}{N!\Lambda^{3N}} \; .
 \end{equation}
 Note that all terms in the sum have the same number of factors $\Lambda^3$ and $q^{int}$. I take these out and define $Q'_{tot}$ through
 \begin{equation}
 Q_{tot} = \frac{(q^{int})^M}{\Lambda^{3M}} Q'_{tot} \; .
 \end{equation}
Note that $Q'_{tot}$ does not depend on $h$. 
Now, the probability to find $N$ particles in the system is:
\begin{equation}
P(N)=\frac{V_0^{M-N}\int d{\bf r}^N\; \exp\left(-\beta U(r^N)\right)}{Q'_{tot}(M-N)!N!}\; .
\end{equation}
Dividing numerator and denominator by $V_0^M$, and using the fact that for $M\gg N$, $(M-N)!/M!\approx M^{-N}$, we can write
 \begin{equation}
P(N)=\frac{\rho_0^N/N! \int d{\bf r}^N\; \exp\left(-\beta U({\bf r}^N)\right)}{\sum_{N=0}^\infty \rho_0^N/N! \int d{\bf r}^N\; \exp\left(-\beta U({\bf r}^N)\right)} \; .
\end{equation}
Note that in this expression, the number of particles in the system is controlled by $\rho_0$, the number density in the reservoir. $h$ has disappeared, as it should.
The concept of a reservoir that contains an ideal gas of molecules that have normal intra-molecular interactions is extremely useful in practice, in particular for flexible molecules.  In a grand canonical simulation, we prepare thermally equilibrated conformations of the molecules that we wish to insert. We can then attempt to insert one of these particles into the system that already contains $N$ particles. The probability of insertion is determined by:
\[
P_{acc}=\min\left\{1,\frac{\rho_0 V}{N+1} e^{-\beta \Delta U} \right\}\;,
\]
where $\Delta U$ measures the interaction of the added particle with the $N$ particles already present, but {\em not} the intramolecular interactions of the added particle.  Note (again) that the crucial control parameter is $\rho_0$, the hypothetical but purely classical density in the ideal reservoir.

Occasionally, the role of Planck constant can be a bit more subtle. We have assumed that the quantised intramolecular motions of the particles do not depend on the inter-molecular interactions. At high densities, this approximation may break down. Usually, this means that the distinction between inter and intra-molecular degrees of freedom becomes  meaningless. However, if we insist on keeping the distinction, then we must account for the shift of intra-molecular energy levels due to inter-molecular interactions. In that case, $h$ necessarily enters into the description.

\subsection{And there is more}
Of course, the examples above are but a small subset of the many subtleties that one may run into when performing a simulation. 
My reason for discussing this particular set is that 1) they are important and 2) they illustrate that one should never use a simulation programme as a `black box'.

\section{Methods that seem reasonable\ldots but are not}
In the previous section, I discussed aspects of simulations that are more tricky than they might at first seem to be.
In the present section, I list a few examples of simulations techniques that cannot be used because they are intrinsically flawed.
\subsection{Computing partition functions by MC sampling}
In a Monte Carlo simulation, we compute thermal averages of the type
\begin{equation}\label{eq:Aav}
\langle A\rangle = \frac{\int d{\bf r}^N \; \exp\left(-\beta U({\bf r}^N)\right) A({\bf r}^N)}{\int d{\bf r}^N \; \exp\left(-\beta U({\bf r}^N)\right)}\; .
\end{equation}
In this expression, $A$ stands for any `mechanical' property of the system (e.g. the potential energy or the virial). 
What we cannot compute in this way are `thermal' properties, such as the free energy of the entropy.  Let us take the free energy as an example.
The relation between the Helmholtz free energy $F$ and the partition function $Q$ of the system is
\begin{equation}
F=-k_BT \ln Q \;.
\end{equation}
In what follows, I will focus on the configurational part of the partition function $Z$, as the remaining contributions to $Q$ can usually be computed analytically.
\begin{equation}\label{eq:Z}
Z=-k_BT\ln\left(\frac{1}{N!}\int d{\bf r}^N \; \exp\left(-\beta U({\bf r}^N)\right)\right)
\end{equation}
Clearly, the integral on the righthand side is not a ratio of two integrals, as in Eqn.~\ref{eq:Aav}. However, we can write
\begin{equation}
\int d{\bf r}^N \; e^{-\beta U({\bf r}^N)}= \frac{V^N\int d{\bf r}^N \; \exp\left(-\beta U({\bf r}^N)\right)}{\int d{\bf r}^N \; \exp\left(-\beta U({\bf r}^N)\right)\exp\left(+\beta U({\bf r}^N)\right)}\; .
\end{equation}
We can rewrite this as
\begin{equation}
\frac{V^N}{\langle \exp\left(+\beta U\right)\rangle}
\end{equation}
and hence it would seem that we {\em can} express the partition function in terms of s thermal average that can be samples.
However, the method will not work (except for an ideal gas, or for similarly trivial systems). The reason is that the most important contributions are those for which $e^{+\beta U}$ is very large, whilst the Boltzmann weight ($\exp\left(-\beta U\right)$) is very small - these parts of configuration space will therefore never be sampled. For systems with hard-core interactions, the situation is even more extreme: an important contribution to the average comes for parts of configuration space where $\exp\left(-\beta U\right)=0$ and $\exp\left(+\beta U\right)=\infty$. In other words, there is no free lunch: $Z$ (and therefore $F$) cannot be determined by normal Monte Carlo sampling. That is why dedicated simulation techniques are needed to compute free energies.  
\subsubsection{\em Particle removal}
The Widom `particle-insertion' method is used extensively to compute (excess) chemical potentials of the components of a moderately dense fluid.
Below I briefly show the derivation of Widom's expression. However, an almost identical derivation leads to an expression that relates the chemical potential to the Boltzmann factor associated with particle removal. That expression is useless and, in the limit of hard-core interactions, even wrong. 

Consider the definition of the chemical potential $\mu_\alpha$ of
a species $\alpha$. From thermodynamics we know that $\mu$  can be
defined as:
\begin{equation}
\label{eqn:mu} \mu =  \left( \frac{\partial F}{\partial
N}\right)_{VT} \; ,
\end{equation}
where $F$ is the Helmholtz free energy of a system of $N$
particles in a volume $V$, at temperature $T$. For convenience, I
focus on a one-component system.  If we express the Helmholtz free
energy of an $N$-particle system in terms of the partition
function $Q(N,V,T)$, then it is obvious that for sufficiently
large $N$ the chemical potential is given by: $\mu = -k_BT
\ln\bigl(Q(N+1,V,T)/Q(N,V,T)\bigr)$. For instance, for an ideal gas of
atoms,
\begin{equation}
\mu^{\rm id}=-k_BT\ln\left[\frac{V}{\Lambda^{3}(N+1)}\right]\;.
\end{equation}
The excess chemical potential of a system is defined as
\begin{eqnarray}
\mu^{\rm ex}(N/V,T)&=&\mu(N/V,T)-\mu^{\rm id}(N/V,T)\nonumber\\
&=& -k_BT\ln\left\{\frac{\int d{\bf r}^{N+1}\exp\left[-\beta
U({\bf r}^{N+1})\right]}{V\int d{\bf r}^{N}\exp\left[-\beta U({\bf
r}^N)\right]}\right\}\;.
\end{eqnarray}
We now separate the potential energy of the $(N+1)$-particle system
into the potential energy function of the $N$-particle system,
$U({\bf r}^N)$, and the interaction energy of the $(N+1)$-th
particle with the rest: $\Delta U_{N,N+1} \equiv U({\bf
r}^{N+1})-U({\bf r}^N)$. Using this separation, we can write
$\mu_{\rm ex}$ as:

\begin{eqnarray}
\mu^{\rm ex}(N/V,T)&=&-k_BT\ln\left\{\frac{\int d{\bf r}_{N+1}\int
d{\bf r}^{N}\exp\left[-\beta U({\bf r}^{N})\right]\exp\left[-\beta\Delta
U_{N,N+1}\right]}{V\int d{\bf
r}^{N}\exp\left[-\beta U({\bf r}^N)\right]}\right\} \nonumber\\
&=& -k_BT\ln\left\{\frac{\int d{\bf
r}_{N+1}\left\langle\exp(-\beta\Delta
U_{N,N+1})\right\rangle_N}{V}\right\}\;,
\end{eqnarray}
where $\left\langle\cdots\right\rangle_N$ denotes canonical
ensemble averaging over the configuration space of the
$N$-particle system. In other words, the excess chemical potential
is related to the average of $\left\langle\exp(-\beta\Delta
U_{N,N+1})\right\rangle_N$ over all possible positions of particle
$N+1$.  In a translationally invariant system, such as a liquid,
this average does not depend on the position of the addition
particle, and we can simply write
\begin{equation}
\label{eqn:Widom2} \mu^{\rm ex}(N/V,T)=-k_BT\ln
\left\langle\exp(-\beta\Delta U_{N,N+1})\right\rangle_N\;.
\end{equation}
The Widom method to determine the excess chemical potential is
often called the ``particle-insertion method" because it relates
the excess chemical potential to the average of the Boltzmann
factor $\exp(-\beta\Delta U_{N,N+1})$, associated with the random
insertion of an additional particle in a system where $N$
particles are already present. 

To arrive at the (incorrect) `particle-removal' expression we simply write
\begin{eqnarray}
\mu^{\rm ex}(N/V,T)&=&\mu(N/V,T)-\mu^{\rm id}(N/V,T)\nonumber\\
&=& +k_BT\ln\left\{V\frac{\int d{\bf r}^{N}\exp\left[-\beta
U({\bf r}^{N})\right]}{\int d{\bf r}^{N+1}\exp\left[-\beta U({\bf
r}^N+1)\right]}\right\}\;.
\end{eqnarray}
Following the same steps as above, we then find:
\begin{equation}
\label{eqn:WidomRemoval} \mu^{\rm ex}(N/V,T)=k_BT\ln
\left\langle\exp(+\beta\Delta U_{N,N+1})\right\rangle_{N+1}\;.
\end{equation}
If the potential energy of the system is bounded from above (and, of course, from below) then this expression is not wrong, but simply rather inefficient.
The reason why it is inefficient is the same as in the case of sampling partition functions: the most important contribution come from regions of configuration space that are rarely sampled. If the potential energy function is not bounded from above (as is the case for hard-core interaction), Eq.~(\ref{eqn:WidomRemoval}) simply yields the wrong answer. Having said that, particle-insertion and particle-removal can be fruitfully combined in the so-called `overlapping-distribution' method~\cite{shi831,ben761}.

\subsection{Using grand-canonical simulations for crystalline solids}
But even the particle insertion method may occasionally yield nonsensical results. This happens when applying the method to a system at high densities. As an example, let us consider a crystalline material. During a simulation of a perfect crystal containing $N$ particles, we can carry out random particle insertions and we can compute the average Boltzmann factor associated with such insertions. The method appears to work, in the sense that it gives an answer. However, that answer is {\em not} the excess chemical potential of atoms in a crystal lattice. Rather, it is the excess chemical potential of {\em interstitial} particles. The reason why the method does not yield the expected answer is that in a real crystal, there are always some vacancies. The vacancy concentration is low (ranging from around 1:10$^4$ at melting, to vanishingly values at low temperatures). Yet, for particle insertion, these vacancies are absolutely crucial: the contribution to the average of $\exp\left(-\beta\Delta U\right)$ of a single insertion in a vacancy far outweighs all the contributions due to insertion in interstitial positions. Using a combination of particle insertion and particle removal~\cite{Pronk} we can obtain information about the equilibrium concentration of vacancies and interstitials. 

Because there are so few vacancies,  a naive grand-canonical simulation of crystal will be inefficient. As the vacancy concentration is low very large systems are needed to create enough vacancies where particles can be added to the system. Of course, one can perform biased grand-canonical simulations to alleviate this problem, but I am unaware of any attempts to do so~\footnote{While writing this manuscript, a preprint by Wilding and Sollich appeared that describes a closely related problem, namely the determination of the chemical potential of a crystalline solid by `ghost'-particle insertion ~\cite{WildingSollich}.}. 

Because the concentration of vacancies is so low in most (but not all~\cite{FilionCubes,FrenkelCubes}) crystals, one subtlety is often overlooked. The point is that, because of the presence of vacancies, the number of particles is not the same as the number of lattice sites (I assume, for simplicity, a one-component Bravais lattice). The free energy of the system is then a function of the number of particles {\em and} the number of lattice sites. We can then write the variation of the free energy at constant temperature as
\begin{equation}
dF = -VdP+\mu dN + \mu_c dN_c \; ,
\end{equation}
where $N_c$ denotes the number of lattice sites and $\mu_c$ the corresponding `chemical potential'. I use quotation marks here because $\mu_c$ is not a true chemical potential: in equilibrium, the number of lattice sites is such that the free energy is a minimum and hence $\mu_c$ must be zero. However, in a simulation where we constrain $N$ and $N_c$ to be the same, $\mu_c$ is definitely not zero and incorrect predictions of the phase diagram result if this is not taken into account. There are only few examples where the analysis of the chemical potential associated with lattice sites has been taken into account correctly~\cite{Mladek,FilionCubes,WildingSollich}.  
\begin{figure}[h!]
\begin{center}
\includegraphics[width=0.8\textwidth]{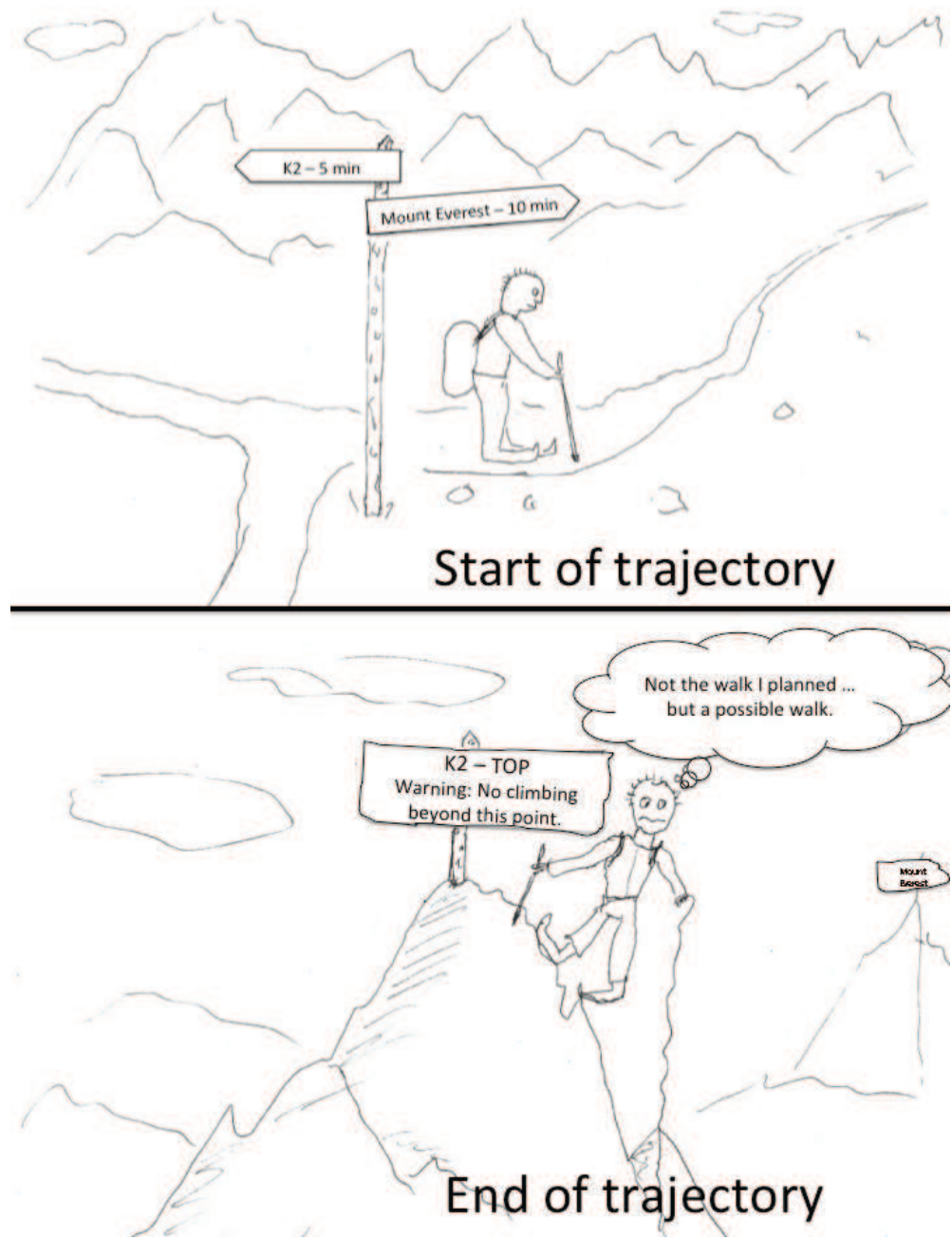} 
\caption{A walker starts on a trajectory with well defined initial conditions (top). However, due to an accumulation of small errors, the walker follows a very different trajectory. Yet that trajectory is (close to) another possible trajectory. }
\label{fig:Lyapunov}
\end{center}
\end{figure} 
\section{Myths and misconceptions}
I end this paper with a very brief summary some `urban legends' in the simulation field. I should apologise to the reader that what I express are my personal views -- not everybody will agree.
\subsection{New algorithms are better than old algorithms}
In the beginning of this paper I pointed out that in a number of specific cases the computational gains made by novel algorithms far outweigh the increase in computing power expressed by Moore's law. However, the number of such algorithms is very small. This is certainly true if we avoid double counting: successful algorithms tend to be re-invented all the time (with minor modifications) and then presented by the unsuspecting authors as new (for example, Simpson's rule has recently been reinvented in a paper that has attracted over 150 citations~\cite{SimpsonTai}).

Although I have no accurate statistics, it seems a reasonable to assume that the number of new algorithms that is produced every year keeps on growing. Unfortunately, these `new' algorithms are often not compared at all with existing methods, or the comparison is performed such that the odds are stacked very much in favour of the new method. 

Emphatically, I am not arguing against new algorithms -- quite the opposite is true and, in fact, over the past few decades we have seen absolutely amazing novel techniques being introduced in the area of computer simulation. I am only warning against the `not invented here' syndrome. 
\subsection{Molecular dynamics simulations predict the time evolution of a many-body system}
This view is more common outside the field than inside. However, the outside community is much larger than the tribe of simulators. 
The reason why Molecular Dynamics does not predict the time evolution of a many-body system is that the dynamics of virtually all relevant many-body systems is chaotic. 
This means that a tiny error of the phase-space trajectory of the system will grow exponentially in time, such that in a very short time the numerical trajectory will bear no resemblance to the `true' trajectory with the same initial conditions. Using more accurate algorithms can postpone this so-called Lyapunov instability. However, doubling the accuracy of the simulation simply shifts the onset by a time $\tau$. Doubling it again, will gain you another time interval $\tau$. The point is that the computational effort required to suppress the Lyapunov instability below a certain preset level scales exponentially with simulation time: in practice, no simulation worthy of the name is short  enough to avoid the Lyapunov instability. It would seem that this is a serious problem: if Molecular Dynamics does not predict the time evolution of the system, then what is the use of the method?

The reason why Molecular Dynamics simulations are still widely used is that (most likely) the trajectories that are generated in a simulation `post-dict' rather than `pre-dict' a possible real trajectory of the system. In particular, it can be shown that the widely used Verlet algorithm generates trajectories that are a solution to the discretised Lagrangian equations of motion. This means that the algorithm generates a trajectory that starts at the same point as a real trajectory, ends at the same point as that trajectory and takes the same time to move from the starting point to the end point.  What Molecular Dynamics generates is a good approximation to a {\em possible} trajectory, not to the trajectory with the same initial conditions. The distinction between the two types of trajectories is illustrated in Fig.~\ref{fig:Lyapunov}. Typically, the time between the beginning and end points of a trajectory is much longer than the time it takes for the Lyapunov instability to appear. Therefore, the trajectory that we generate need not be close to any real trajectory. However, if we consider the limit where the time-step goes to zero, then the discretised Lagrangian approaches the full Lagrangian and the discretised path (presumably) approaches the real path. I should stress that this result holds for the Verlet algorithm and, presumably, for a number of Verlet-style algorithms. However, it is not obvious (to me) that the Molecular Dynamics trajectories generated in an event-driven algorithm with hard collisions are necessarily close to a real trajectory for long times. 
If we can believe Molecular Dynamics simulations (and I believe that we can in a `statistical' sense) then there are several excellent reasons to use the technique. The first is, obviously, that Molecular Dynamics probes the dynamics of the system under study and can therefore be used to compute transport coefficients, something that a Monte Carlo simulation cannot do. Another practical reason for using Molecular Dynamics is that it can be (and, more importantly, has been) parallellized efficiently. Hence, for large systems, Molecular Dynamics is almost always the method of choice.

\subsection{Acceptance of Monte Carlo moves should be around 50 \%}
This rule-of-thumb is old and nowhere properly justified. There are many examples where it appears that the optimal acceptance of Monte Carlo moves should be substantially lower (certainly for hard core systems). Also, in the mathematical literature~\cite{rob971} it is argued (but in a different context) that the optimal acceptance of Metropolis-style Monte Carlo moves is probably closer to 25 \% (23.4 \% has been shown to be optimal for a specific -- but not a statistical mechanical -- example). 

\begin{figure}[h!]
\begin{center}
\includegraphics[width=0.8\textwidth]{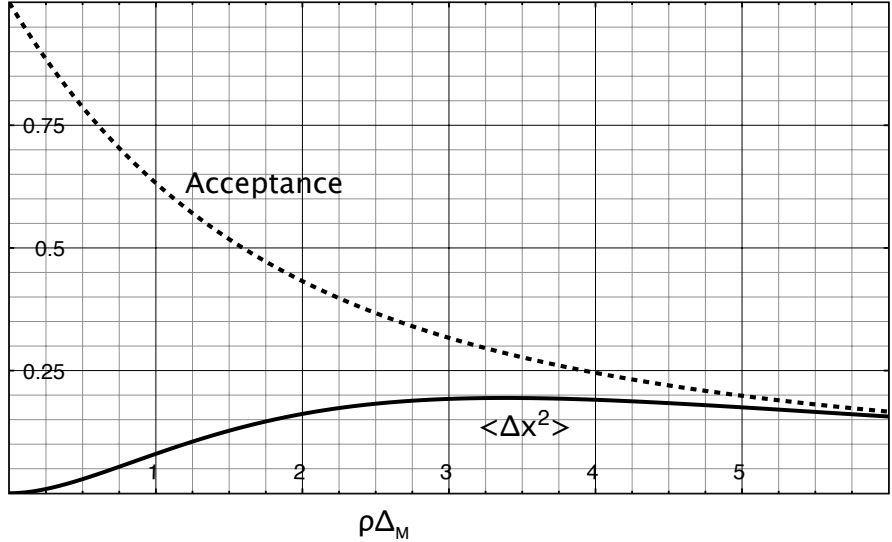} 
\caption{Relation between acceptance of Monte Carlo moves and sampling efficiency. In the example discussed in the text, the most efficient algorithm is the one that diffuses most rapidly through configuration space. Solid curve: mean-squared displacement per move, as a function of maximum trial displacement. Dotted curve shows the corresponding acceptance probability. The figure shows that the largest mean-square displacement occurs for an acceptance ratio of approximately 28 \% -- i.e. well below 50 \%. }
\label{fig:acceptance}
\end{center}
\end{figure}

A very simple example can be used to illustrate this point. Consider a hard particle in an ideal gas. The number density of the ideal gas is denoted by $\rho$. If we move the hard particle over a distance $\Delta x$, it sweeps out a volume proportional to $\Delta x$. To make life easy, I will consider a one-dimensional example. In that case the volume swept out by the hard particle is equal to $\Delta x$ (I will assume that the particle itself is much larger than $\Delta x$). The probability that a Monte Carlo  move over a distance $\Delta x$ will be accepted is given by the probability that there will be no ideal gas particles in the volume swept out by the particle:
\begin{equation}
P_{acc}(\Delta x)=\exp\left(-\rho|\Delta x|\right) \;.
\end{equation}
If the trial displacement is $\Delta x$, then the average {\em accepted} displacement is $\Delta x P_{acc}(\Delta x) +\left[1-P_{acc}(\Delta x)\right]\times 0$ and the corresponding mean-squared displacement is
\begin{equation}
\langle x^2(\Delta x)\rangle = (\Delta x)^2\exp\left(-\rho|\Delta x|\right)\;.
\end{equation}
Of course, this system is trivial because we know that if we sample all possible configurations, the large particle will be distributed uniformly in the volume occupied by the ideal gas. Again, in this case, it is plausible that the most efficient Monte Carlo algorithm is simply the one that makes the large particle `diffuse' most rapidly through configuration space.
Now consider that we generate trial moves such that $\Delta x$ is uniformly distributed between $+\Delta_M$ and $-\Delta_M$, where $\Delta_M$ denotes the magnitude of the maximum displacement. The diffusion of the large particle will be fastest for that value of $\Delta_M$ that yields the largest mean-squared displacement. We can compute this mean-squared displacement analytically:
\begin{equation}
\langle x^2\rangle_{\Delta_M}=2\Delta_M^2 \frac{1-\exp\left(-\rho\Delta_M\right)\left[1+\rho\Delta_M+(\rho\Delta_M)^2/2\right]}{(\rho\Delta_M)^3}\; .
\end{equation} 
As can be seen in Fig.~\ref{fig:acceptance}, the maximum value of $\langle x^2\rangle_{\Delta_M}$ corresponds to an average acceptance $[1-\exp(-\rho\Delta_M)]/(\rho\Delta_M)$ that is around 28 \% (i.e. considerably less than 50\%).

Does it matter? Probably not very much: the very fact that there is an `optimal' acceptance ratio means that the efficiency is fairy flat near this optimum. Still, 25 \% is rather different form 50 \%. When in doubt, check it out.
\subsection{Periodic boundaries are boundaries}
This is a point that I already mentioned above but I repeat it here: in normal (equilibrium) Monte Carlo or Molecular Dynamics simulations, the origin of the periodic box can be chosen wherever we like and we can remap the periodic boxes at any time during our simulation. Hence, the `boundary' of the periodic box is not a physical divide. Again, as said before, this situation changes for some non-equilibrium situations, e.g. those that use Lees-Edwards sliding boundary conditions~\cite{LeesEdwards}.

There is another situation where the boundary of the simulation box has (a bit) more meaning, namely in the simulation of ionic systems. Due to the long-range nature of the Coulomb forces, the properties of the system depend on the boundary conditions at infinity. This is not very mysterious: if we have a flat capacitor with a homogeneous polarised medium between the plates, then the system experiences a depolarisation field, no matter how far away the plates are. However, if the capacitor is short-circuited, there is no depolarisation field. 

Something similar happens in a simulation of a system of charged particles: we have to specify the boundary conditions (or, more precisely, the dielectric permittivity) at infinity. If we choose a finite dielectric constant at the boundaries (`finite' in this context means: anything except infinite) then the internal energy of the system that we simulate has a polarisation contribution that depends quadratically on the dipole moment of the simulation box. If we study a dipolar fluid, than the dipole moment of the simulation box does not depend on our choice of the origin of the box. However, if we study a fluid of charged particles, then the choice of the boundaries does matter. Again, it is nothing mysterious: some crystal faces of ionic crystals are charged (even though the crystal is globally neutral). The depolarisation field in this crystal will depend on where we choose the boundary (at a positively  or at a negatively charged layer), no matter how large the crystal.

\subsection{Free energy landscapes are unique and meaningful}
If the potential energy $U$ of a system is known as a function of the particles coordinates ${\bf r}^N$, then it is meaningful to speak about {\em the} energy `landscape' of the system: the potential energy of an $N$ particle system in $d$ dimensions will have minima, saddle points and maxima in the $dN$ dimensional space spanned by the cartesian  coordinates of the particles.

In contrast, it is not meaningful to speak of {\em the} free energy landscape: there are many. In general, free energies appear when we wish to describe the state of the system by a smaller number of coordinates than $dN$. Of course, using a smaller number of coordinates to characterise the system will result in a loss of information. Its advantage is that it may yield more physical insight -- in particular if we use physically meaningful observables as our coordinates. Examples are: the total dipole moment of a system, the number of particles in a crystalline cluster or the radius of gyration of a polymer. Typically, we would like to know the probability to find the system within a certain range of these new variables.  This probability (density) is related to the original Boltzmann distribution. In what follows, I will consider the case of a single coordinate $Q$ (often referred to as the `order parameter'). The generalisation to a set of different $Q$'s is straightforward. Note that $Q$ may be a complicated, and often non-linear function of the original particle coordinates  ${\bf r}^N$.
The probability to find the system in a state between $Q$ and $Q+dQ$ is $p(Q)dQ$, where $p(Q)$ is given by
\begin{equation}
p(Q) =
\frac{
\int d {\bf r}^N \exp(-\beta {\cal U}_0)\delta[Q-Q({\bf r}^N)]
}{
Z
}\;,
\end{equation}    
where $Z$, the configurational part of the partition function,  is given by
\begin{equation}\label{eq:ZiWi}
Z_\equiv
\int d{\bf r}^N \exp\left(-\beta U({\bf r}^N)\right)\; .
\end{equation} 
We now {\em define} the free energy `landscape' $F(Q)$ as 
\begin{equation}
F(Q)=-k_{\rm B}T\ln P(Q)
\end{equation}
which implies that
\begin{equation}
P(Q)=\exp\left(-\beta F(Q)\right)\;. 
\end{equation}
Hence, the $F(Q)$ plays the same role in the new coordinate system as $U({\bf r}^N)$ in the original $dN$ dimensional system. 
The free energy $F(Q)$ may exhibit maxima and minima and, in the case of higher dimensional free-energy landscapes, free-energy minima are separated by saddle points. These minima and saddle points play an important role when we are interested in processes  that change the value of $Q$, e.g. a phase transition, or the collapse of a polymer. The natural assumption is that the rate of such a process will be proportional to the value of $\exp\left(-\beta F(Q)\right)$ at the top of the free-energy barrier separating initial and final states. However, whilst, for reasonable choose of $Q$, this statement is often approximately correct, it is misleading because the height of the free energy barrier depends on the choice of the coordinates. To see this, consider another coordinate $Q'$ that measures the progress of the same transformation from initial to final state. For simplicity, I assume that I have a one-dimensional free-energy landscape and that $Q'$ is a function of $Q$ only: $Q'(Q)$. Then it is easy to show that
\begin{equation}
\exp\left(-\beta F'(Q')\right)=\exp\left(-\beta F(Q)\right)\left|\frac{\partial Q}{\partial Q'}\right| \;.
\end{equation}
The derivative on the right-hand side is the Jacobian associated with the coordinate transformation from $Q$ to $Q'$. If $Q'$ is a linear function of $Q$, then the Jacobian is harmless: it simply adds a constant to the free energy and hence does not affect barrier heights. However, in the more general case where $Q'$ is a non-linear function of $Q$, the choice of the functional form affects the magnitude of the barrier height. Let us consider an extreme (and unphysical) examples to illustrate this. We choose:
\begin{equation}\label{eq:JacobianTransformation}
Q'=\int_{Q_0}^Q dq\; e^{-\beta F(q)}.
\end{equation}
Then
\begin{equation}
\exp\left(-\beta F'(Q')\right)=\exp\left(-\beta F(Q)\right)\exp\left(+\beta F(Q)\right)=1 \;.
\end{equation}
In other words: with this choice of coordinates, the barrier has disappeared altogether.
In general, the choice of different order parameters will not affect the free energy barrier as drastically as this, but the fact remains: the height of the free energy barrier depends on the choice of the order parameter(s) $Q$. 

Of course, this dependence of the free-energy barrier on the choice of the order parameter is only an artefact of our description. It cannot have physically observable consequences. And, indeed, it does not. The observable quantity is the rate at which the initial state transforms into the final state (folding rate, nucleation rate etc.) and that rate is independent of the choice of $Q$. 

It is instructive to consider a simple example: the diffusive crossing of a one-di\-men\-si\-onal barrier of height $\epsilon$ and with a flat top of width $W$
\begin{equation}
F(Q)= \epsilon \left[\theta(Q)-\theta(Q+W)\right]\; ,
\end{equation}
where $\theta(x)$ is the Heaviside step function. We assume that the diffusion constant at the top of the barrier has a constant value $D$. Moreover, we assume that, initially, the density  is $c$ on the reactant side, and zero on the product side. Then the initial, steady-state flux across the barrier is given by:
\begin{equation}
J=\frac{cD}{W} \exp\left(-\beta F(Q)\right)\; .
\end{equation}
Now we apply our transformation Eq.~(\ref{eq:JacobianTransformation}). In the region of the barrier, we then get:
\[
Q'=Q \exp\left(-\beta \epsilon\right).
\]
This transformation will completely eliminate the barrier. However, it should not have an effect to the reactive flux. Indeed, it does not, as we can easily verify: the width of the barrier in the new coordinates is
\[
W'=W \exp\left(-\beta \epsilon\right)\;,
\]
and the diffusion constant becomes
\[
D'=D \exp\left(-2\beta \epsilon\right) \;.
\]
Combining all terms, we get:
\begin{equation}
J'=\frac{cD'}{W'}\times 1 = \frac{cD}{W} \exp\left(-\beta \epsilon\right)\;,
\end{equation}
which is identical to the original flux, as it should be. 

\begin{figure}[h!]
\begin{center}
\includegraphics[width=0.8\textwidth]{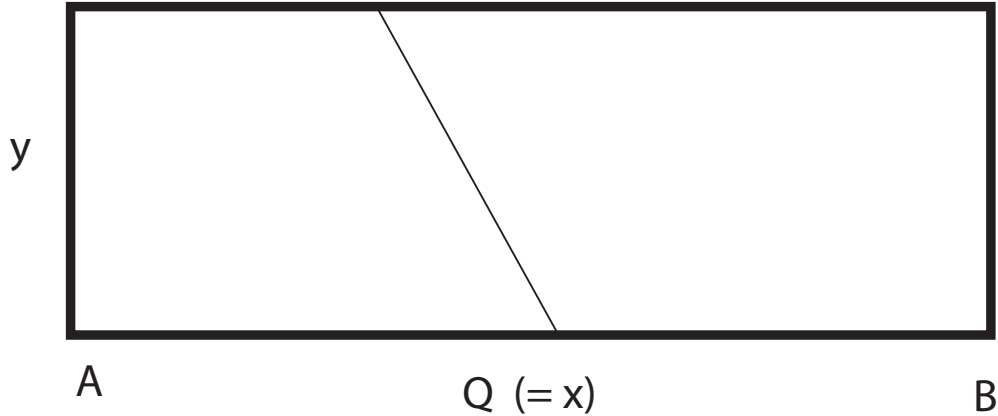} 
\caption{System with an infinitely thin, impenetrable barrier, oriented obliquely with respect to the `order parameter' $Q$. In this system, the free energy as a function of $Q$ is flat, yet there is no physical path from $A$ to $B$.}
\label{fig:nopath}
\end{center}
\end{figure}

There is another problem with free energy barriers: the fact that a coordinate $Q$ joins two minima in a free energy landscape, does not necessarily imply that a physical path exists. Again, we consider a simple example: a two dimensional system of a particle in a rectangular box. The box is cut obliquely by an infinitely thin, but infinitely high, energy barrier (see Fig.~\ref{fig:nopath}). 
We now choose the $x$ coordinate as our order parameter $Q$. The free energy is computed by integrating the Boltzmann factor over $y$ at constant $Q$. The fact that the Boltzmann factor vanishes on one point along a line of constant $Q$ makes no difference to the integral. Hence, $F(Q)$ is constant. The path between $A$ (on the left) and $B$ (on the right) appears barrier free. Yet, clearly, there is no physical path from $A$ to $B$.

The reader should not conclude from the above discussion that free-energy landscapes are useless concepts: far from it. However: 
\begin{enumerate}
\item The construction of the coordinates $Q$ is a matter of choice that should be guided by physical insight or intuition
\item The height of free energy barriers depends on the choice of these coordinates
\item The rate of `barrier-crossing' processes does {\em not} depend on the choice of the coordinates
\item The fact that there is no free-energy barrier separating two states does not necessarily imply that there is an easy physical bath between these states. 
\end{enumerate}

\subsection{And there is more\ldots}
As the field of numerical simulation expands, new techniques are invented and old techniques are rediscovered. But not all that is new is true.
There are several examples of methods that had been discredited in the past that have been given  a new identity and thereby a new lease on life.
Usually, these methods disappear after some time, only to re-emerge again later. The fight against myths and misconceptions never ends (including the fight against fallacies that I have unknowingly propagated in this paper). 

\acknowledgments
I am grateful to Wilson Poon for suggesting the topic of this paper. What I wrote is necessarily very incomplete, but I enjoyed writing it.
I thank Patrick Varilly for a critical reading of this manuscript and for his many helpful suggestions. I should stress that all remaining errors and misconceptions are mine alone.
This work was supported by the European Research Council (ERC) Advanced Grant 227758, the Wolfson Merit Award 2007/R3 of the Royal Society of London and the Engineering and Physical Sciences Research Council (EPSRC) Programme Grant EP/I001352/1.

\end{document}